\documentclass[aps,prb,twocolumn,showpacs]{revtex4}
\pdfoutput=1
\usepackage{amsmath}  
\usepackage{graphicx} 
\usepackage{subfigure}
\begin{document}
\title{Computation in Classical Mechanics}
\author{Todd Timberlake}
\affiliation{Department of Physics, Astronomy, \& Geology, Berry
College, Mount Berry, GA  30149}
\email{ttimberlake@berry.edu}

\author{Javier E. Hasbun}
\affiliation{Department of Physics, University of West Georgia, Carrollton, GA  30117}%
\homepage{http://www.westga.edu/~jhasbun}
\email{jhasbun@westga.edu}

\date{\today}

\begin{abstract}
There is a growing consensus that physics majors need to learn
computational skills, but many departments are still devoid of
computation in their physics curriculum.  Some departments may
lack the resources or commitment to create a dedicated course or
program in computational physics.  One way around this difficulty
is to include computation in a standard upper-level physics
course. An intermediate classical mechanics course is particularly
well suited for including computation.  We discuss the ways 
we have used computation in our classical mechanics courses,
focusing on how computational work can improve students'
understanding of physics as well as their computational skills.  We
present examples of computational problems that
serve these two purposes.  In addition, we provide information about resources
for instructors who would like to include computation in their
courses.
\end{abstract}
\maketitle

\section{Introduction}

The primary purpose of this article is to suggest a method for incorporating
computation into upper-level classical mechanics courses.  There is an emerging
consensus in the physics community that computational skills are important for
physicists,\cite{Grayson} but too often there is little or no computation
included in the physics curriculum.\cite{Fuller}  Probably the best way to
strengthen the computational component of the physics curriculum is to
incorporate computation into all components of the curriculum.  Several
institutions have developed large-scale computational physics programs,\cite{programs}
which typically include at least one dedicated computational physics course as
well as computational components in other upper-level courses.  Other
institutions offer a single computational physics course along with an otherwise
traditional curriculum.  With a variety of good computational physics textbooks\cite{comptexts}
available it may appear that there is no reason not to offer at
least one course in computational physics.  Unfortunately, the reality is that
there are constraints that may prevent some departments from offering such a course.
These constraints may involve the number of physics faculty, the computational
background of the physics faculty, low student enrollment in physics courses,
or even political resistance to curricular changes.  For faculty who face these
constraints but desire to include some computation in the physics curriculum,
the best approach may be to incorporate computation into existing
courses. \cite{Roos}  Even departments that have a computational physics
course may be looking for ways to increase the role of computation in the
physics curriculum.  We believe that the intermediate classical mechanics
course, typically taken by students in their sophomore or junior year, is
ideally suited for incorporating computation.

Computation can contribute to a variety of learning goals, and
different approaches to incorporating computation into physics
courses will likely emphasize different goals.  One goal of
including computation in physics courses is simply to improve
students' learning of physics concepts. Computer
visualizations and interactive simulations are particularly useful in this regard.  
Computation can also
open the door to new and important topics such as nonlinear dynamics and chaos.  
Including computation in physics courses can
also increase physics students' familiarity with widely-used
computational tools and help students see how these tools can be
applied to solving physics problems.  Finally, computation can be
used to introduce students to important numerical algorithms.  In
many cases these algorithms can provide insight into important
physics concepts, in addition to serving as tools for carrying out
computations.

Although all of these learning goals can be achieved through a
dedicated course in computational physics, as mentioned above this
may not be a practical approach in all departments.  Incorporating
computation into a standard course may be an effective way to
introduce computation into the physics curriculum, perhaps as a
temporary solution until a dedicated computational course can be
added.  Computation can be included in a standard course even if
the students have no computational background (although, of
course, more can be done if the students have had a course in
programming or computational physics). We feel that the standard
(sophomore/junior level) intermediate classical mechanics course
is well suited for including computation. Students in classical
mechanics can benefit tremendously from computer visualizations,
which help them to build intuition about classical dynamics.
Classical mechanics provides an excellent forum for introducing a
variety of important numerical tools such as ODE solvers, root finding, numerical integration, 
numerical linear algebra, etc.  Some
important topics in modern classical mechanics, like chaos, cannot
be effectively taught without computation. Another advantage of the classical mechanics course is
that it is
typically taught early in the upper-level physics curriculum,
often immediately after the introductory sequence.  Introducing
students to computation at this early stage gives them the
opportunity to use their computational skills in later physics and
mathematics courses.

\section{Implementation Issues}

To begin using computation in a classical mechanics course one
must first make choices about which computational tools to use and
how to use them.  Choosing the right computational platform can be
difficult, as each platform has advantages and disadvantages.
Symbolic and numeric mathematics software packages (such as
Mathematica, Maple, and Matlab, each of which has its own
advantages and disadvantages\cite{3M}) offer quick and relatively
easy ways to perform computational work, provide a wide range of
computational tools, and include high-quality visualization tools.
The downside is that these tools can be used with little
understanding of the numerical methods that are being employed.
Furthermore, these software packages can be quite expensive.  The
other main alternative is to have students create programs from
scratch using a standard programming language like Fortran, Java,
or Python.  Students who do their own coding must develop an understanding 
of the models they are
studying as well as the numerical methods that are employed.  This solution is generally inexpensive since free compilers are available for these languages on a
variety of operating systems.  Many
common programming tasks can be carried out with the aid of
freely-available numeric libraries, like Open Source Physics
(OSP)\cite{OSP} and SciPy,\cite{SciPy} to reduce the time that
must be spent creating the programs.  However, in spite of these
libraries there is still a great deal of programming ``overhead''
that must be addressed and instructors whose students have no
programming background may not have sufficient time to teach the
necessary programming skills.  There are some intermediate
solutions available, such as the Easy Java Simulations (EJS)
package,\cite{EJS} which automates many programming tasks and
allows users to focus on the numerical algorithms that are used.
However, EJS does not offer the broad range of functionality that
is available with a standard programming language or a software
package like Mathematica.

We have chosen to use Mathematica and Matlab in our classical
mechanics courses, but we plan to move toward using Java/OSP and
EJS for some topics.  We like Mathematica for its powerful
symbolic mathematics capabilities, and Matlab because it is widely
used in industry.  Both of these packages are tools that students
will likely be able to use in other classes and perhaps throughout
their careers.  Programming in Java with the OSP library provides
an opportunity to share both programs and curricular materials
with the broader physics community and contribute to (as well as
benefit from) a larger project.\cite{Belloni}  EJS provides simple
tools for quickly creating visualizations and simulations. Each of
these options, as well as the others listed above, will be more or
less suited to an individual instructor's specific situation.
Instructors must be guided by the background of their students (Do
they have programming experience?), their own computational
experience (With what platforms is the instructor familiar?), as
well as other practical considerations (Are there funds available
to pay for software licenses?  Will the software be supported by
qualified campus staff?).

The choice of platform (or platforms) should also be guided by how
the instructor intends to use the computational tools in the
course.  In our classical mechanics courses we use computation in
two main ways:  for in-class demonstrations and for student
computational projects.  Typically we will demonstrate computer
solutions to physics problems as part of our in-class lecture. The
code used to construct the solution is made available to students
so that they can examine the code and ``experiment'' with it at
their leisure.  We then follow up these demonstrations by
assigning computational projects that students must complete on
their own or in small groups.  Early in the course these projects
may involve only minor modifications of the code used in the
demonstrations, perhaps to study the same phenomenon in a new
system.  In this way the demonstration code serves as a template
that helps the students complete the computational project.  As
the course proceeds we demand more from our students, expecting
them to construct computational solutions without a template (but
using computational tools they have seen before).  These computational projects
can easily be made the basis of formal writing assignments in which students
must present their analytic and numerical work along with figures,
typeset equations (possibly in /LaTeX), and a thorough discussion of the physics.

\section{Algorithms and Physics}

Because the software packages we use tend to hide the details of
the numerical algorithms they employ, we feel that it is
important to provide students with some explicit instruction on
algorithms.  We focus on simple algorithms, rather than the sophisticated algorithms
employed by the software packages we use, for a few
important computational tasks.  This instruction
enhances students' knowledge of computation because it emphasizes
the fact that computations require \emph{some} algorithms and that
the choice of algorithm can critically affect the success of the
computation.  In addition, teaching students about algorithms can
sometimes lead to a better understanding of important physics
concepts.

\begin{figure}[t]
\begin{center}
\subfigure{\includegraphics[width=1.5in,height=1.5in]{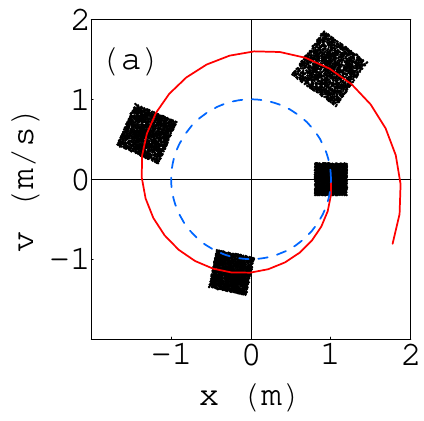}}
\subfigure{\includegraphics[width=1.5in,height=1.5in]{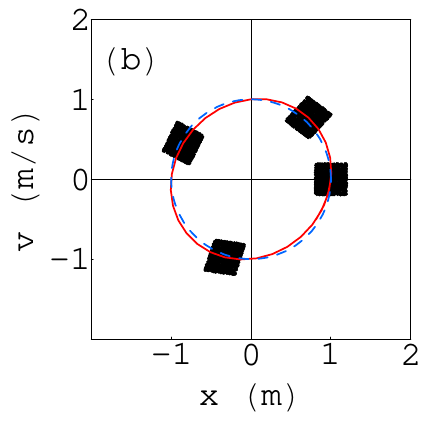}}
\caption{\label{Euler} Comparison of (a) Euler and (b)
Euler-Cromer algorithms applied to the simple harmonic oscillator.
In both cases $k=1$ N/m, $m=1$ kg, $x(0)=1$ m, and $v(0)=0$.  The
dashed line shows the exact orbit in phase space.  The solid line
shows the solution generated by the algorithm.  Also shown is a
cluster of trajectories initially distributed at random in a
square centered on $x = 1$ m, $v=0$.  The locations of these
trajectories resulting from each algorithm are shown at $t = 0$,
1.8, 3.6, and 5.4 s.}
\end{center}
\end{figure}

As an example of how we use algorithms in our course, let us
consider an object of mass $m$ oscillating in one dimension on an
ideal spring with force constant $k$.  The equation of motion for
this simple harmonic oscillator is easy to solve analytically, but we can also take a
numerical approach to the problem.  The simplest algorithm that we
might consider for this purpose is the Euler algorithm:
\begin{equation}
\label{eu}
\begin{array}{lll}
x_{n+1} & = & x_n + v_n \Delta t, \\
v_{n+1} & = & v_n + F_n \Delta t / m, \\
\end{array}
\end{equation}
where $x_n$, $v_n$, and $F_n$ represent the displacement from
equilibrium, the velocity, and the net force at time $t_n = t_0 +
n \Delta t$.  For this system $F_n = -kx_n$.  Although the
software packages we use have their own built-in ODE solvers, we
can also use these packages to implement a simple algorithm like
the Euler algorithm.  The solid red curve in Figure \ref{Euler}(a)
shows the trajectory of the object in phase space (position versus
velocity) generated by the Euler algorithm when $m = 1$ kg, $k=1$
N/m, $x(0)=1$ m, $v(0)=0$, and $\Delta t=0.2$ s.  The dashed blue
curve in Fig. \ref{Euler}(a) shows the exact solution.  It is
clear from this result that the Euler algorithm is unstable since
the trajectory produced by the algorithm spirals continually
outward away from the exact solution.  This also means that the
Euler algorithm does not conserve energy:  it is clear that with
the Euler algorithm the object's maximum displacement and maximum
speed are steadily increasing, so its total mechanical energy must
also be steadily increasing.

The failure of the Euler algorithm to conserve energy is related
to yet another failure:  the failure to preserve phase space
volume.  According to Liouville's Theorem, a conservative system
(such as our simple harmonic oscillator) must preserve the volume
(or area) occupied by an ensemble of trajectories in phase space.
Figure \ref{Euler}(a) shows the locations generated by the Euler
algorithm for a trajectory ensemble at times $t = 0$, 1.8, 3.6,
and 5.4 s.  At $t=0$ the points are randomly distributed within a
square region around $x=1$ m, $v=0$.  As time passes the
cluster of points moves through phase space and the shape of the
region becomes distorted.  However, it is clear from Figure
\ref{Euler}(a) that the size of the region also grows over time,
in clear contradiction to Liouville's Theorem.  In fact, it is
easy to show that the Euler algorithm doesn't preserve phase space
volume.  If we treat Equation \ref{eu} as a two-dimensional map,
then this map will preserve phase space volume if and only if the
Jacobian of the map, defined by
\begin{equation}
\label{jacobian}
J = \left(
    \begin{array}{ll}
    \frac{\partial x_{n+1}}{\partial x_n} & \frac{\partial x_{n+1}}{\partial v_n} \\
    \frac{\partial v_{n+1}}{\partial x_n} & \frac{\partial v_{n+1}}{\partial v_n}
    \end{array}
    \right),
\end{equation}
has determinant equal to one.  A quick calculation will show that
$|J| \neq 1$ for the Euler algorithm, in the case of the simple harmonic oscillator.

These problems with the Euler algorithm can be fixed with a simple
modification, leading to what is known as the Euler-Cromer
algorithm.\cite{Cromer}  This algorithm simply updates the
velocity first, and then uses the new velocity to update the
position.  The equations describing this algorithm are:
\begin{equation}
\label{ec}
\begin{array}{lll}
x_{i+1} & = & x_i + v_{i+1} \Delta t \\
v_{i+1} & = & v_i + F_i \Delta t / m, \\
\end{array}
\end{equation}
where all quantities are defined as for the Euler algorithm above.
 This algorithm is stable and conserves energy, on average, for
oscillatory motion.\cite{Cromer}  We
can quickly illustrate this by applying this new algorithm to our
simple harmonic oscillator. The results are shown in Figure
\ref{Euler}(b).  It is clear that the solution produced by the
algorithm remains close to the exact solution, coinciding with the
exact solution at every quarter period. Similarly, the total
energy oscillates about the correct value with a period equal to
half of the oscillator's period. Furthermore, this algorithm
appears to preserve phase space volume.  This can be easily proved
by showing that the Jacobian for Equation \ref{ec} has determinant
equal to one.

Once these algorithms (and their differences) have been
demonstrated, students should be given a chance to use them.  They
can explore the behavior of the algorithms as the time step is
increased or decreased.  They can compare the results of these
algorithms with the results obtained using  the built-in ODE
solver supplied by the software package.  They can make
modifications to the model by adding drag forces or driving
forces.  Note that students are not restricted to drag forces
linear in the velocity, or sinusoidal driving forces.  In fact, it
is very instructive to consider the case of a quadratic drag force
(a nonlinear system) and compare the results to those obtained
with a linear drag force (a linear system).  If these
non-conservative forces are added then neither algorithm will
preserve phase-space volume, as can be easily seen by finding the
determinant of the Jacobian (and noting that $F_n$ is no longer a
function of $x_n$ only).  Carrying out this calculation using the
Euler-Cromer algorithm makes it clear that for drag forces, where
$\partial F/\partial v < 0$, the determinant of the Jacobian is
less than one indicating that phase space volume will shrink over
time.  This is exactly what one should expect for a dissipative
system.

Other algorithms for common numerical tasks may be worth
discussing in class.  Algorithms for root finding (Newton-Raphson
method, bisection method) and numerical integration (trapezoid
approximation, Simpson's Rule, Monte Carlo methods) are useful and
simple enough to present without absorbing too much class time.
Although class time spent discussing algorithms may take away from time
spent discussing physics, the above example shows that it is
possible to teach important physical concepts along with
algorithms.  As another example, the Newton-Raphson method
illustrates the concept of a stable attractor which is important
for understanding the dynamics of dissipative systems.

\section{Computational Projects}

\begin{figure}[t]
\begin{center}
\includegraphics[width=3.0in,height=3in]{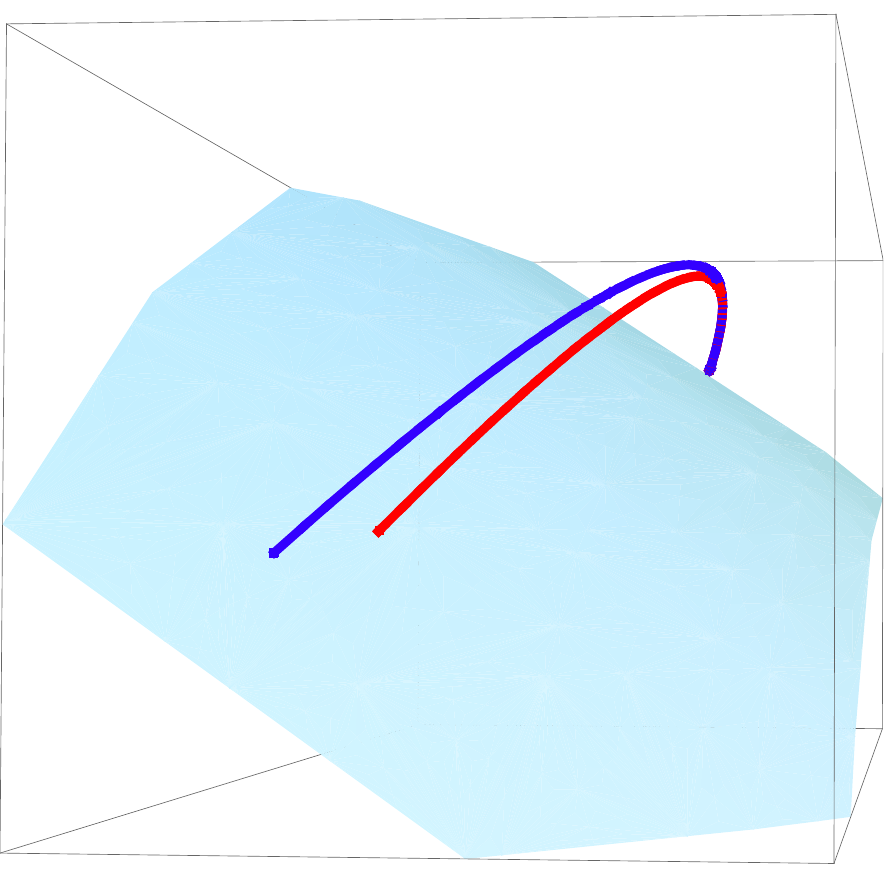}
\caption{\label{proj} Trajectory of a long-range projectile fired due East from Rome, Georgia.  One path is calculated by taking non-inertial forces into account, while the other path ignores these apparent forces.}
\end{center}
\end{figure}

\begin{figure}[t]
\begin{center}
\includegraphics[width=3.0in,height=3in]{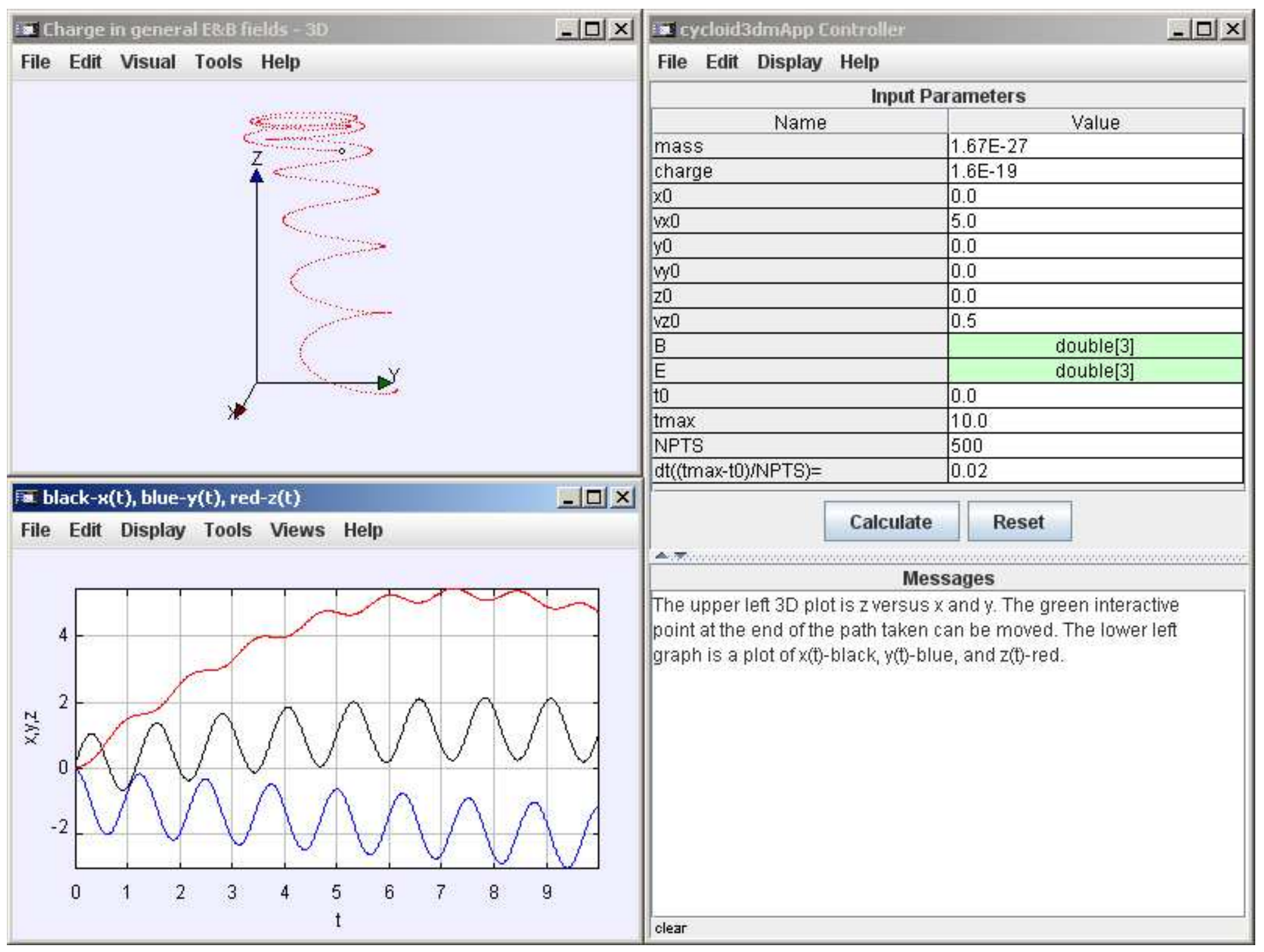}
\caption{\label{charge} An OSP application for visualizing the motion of a charged particle in the presence of
three dimensional constant electric and magnetic fields.}
\end{center}
\end{figure}

Computational assignments should be chosen with care to ensure
that they take full advantage of what computing offers.  One way
to do this is to use computation to break out of the restrictions
imposed by the need for an analytic solution (or approximate
solution).  We have already mentioned the case of the harmonic
oscillator with a quadratic drag force.  Two other examples of
this type are motion in a non-inertial reference frame and motion
of a charged particle in electric and magnetic fields. Traditional
courses tend to focus on cases where analytic solutions are
possible, such as the Eastward deflection of an object that drops
from rest to Earth or the cyclotron motion of an electron in a
magnetic field.  Computation allows students to solve more
realistic problems such as long-range projectile motion on Earth
and the motion of an electron in combined electric and magnetic fields.  Solving such
problems computationally not only allows students to tackle
realistic problems, it also allows students to visualize the
motion in these systems.  Figure \ref{proj} shows two calculated
trajectories, one with non-inertial forces and one without, for a
long-range projectile fired due East from Rome, Georgia.
Visualizing this motion helps students get a sense of the
magnitude of the effect of non-inertial forces. Figure
\ref{charge} shows an OSP application that simulates a charged particle
moving in the presence of constant (but general) three dimensional
electric and magnetic fields. Students can experiment with
the parameters of the calculation without performing a
tedious analytic analysis.

Some topics cannot be taught at all without computation.  The most
obvious example of this is chaos.  Computational solutions of
systems like the driven, damped pendulum can be used to illustrate
important concepts of dissipative chaos such as sensitive
dependence on initial conditions, period doubling bifurcations,
Poincar\'{e} sections, and strange attractors.  Simple iterated
function systems like the logistic map can be used to delve deeper
into dissipative chaos with bifurcation diagrams and Lyapunov
exponents.  While these topics are now included in many
textbooks,\cite{texts, Hasbun} they cannot really be taught
effectively without letting students carry out some
computations.  Once students have mastered the
necessary computational skills, even the more advanced topic of
Hamiltonian chaos becomes accessible through the study of
two-dimensional area-preserving maps.\cite{Timberlake} In fact,
students who have already seen the Euler-Cromer algorithm will be
familiar with some of the properties of area-preserving maps.

Computational projects can also be chosen to introduce
important numerical methods.  Numerical integration can be
introduced in the context of finding the period of a pendulum
undergoing large amplitude oscillations.  Numerical root-finding
techniques can be used to determine the time of flight for a
projectile with air resistance.  Numerical determination of
eigenvalues and eigenvectors can be used in the context of finding
principle axes for rigid-body motion or normal modes of coupled
oscillator systems.

Many textbooks now have computational exercises among their
end-of-the-chapter homework problems, and these can be a valuable
source of ideas for computational projects.\cite{texts}  One of us
(JH) has written a textbook for classical mechanics that
explicitly incorporates computation throughout the book and
includes a wide variety of computational projects.\cite{Hasbun} In
addition, anyone can obtain most of our computational materials
(except solutions to student projects) written in
Matlab\cite{mscript} or Mathematica\cite{mathematica}.   We plan
to update these websites with new materials as they are created.
We also plan to port these materials to Java/OSP or EJS.  These
materials will be made available through the BQ-OSP
database.\cite{bq}  Instructors wishing to include computation in
their physics courses are also encouraged to consult computational
physics texts\cite{comptexts} and the Open Source Physics
website.\cite{OSP}

\section{Summary}

A course in intermediate classical mechanics provides an excellent
opportunity for including computation without requiring the
resources and commitment needed for developing a new computational
course or program.  Many topics that are typically covered in an
intermediate classical mechanics course can benefit from a
computational approach.  The examples discussed in this article
provide just a small sample.  The intermediate classical mechanics
course is also a good place to introduce computation because it
typically comes early in the curriculum.  Ideally this course
would be preceded by a computational physics course and followed
by other upper-level physics courses that include computational
work.  We have found that even if computation is required only in
classical mechanics students will use the computational
skills they have gained to solve problems in a wide variety of
other physics and mathematics courses.  An early exposure to
computational work can also lead to opportunities for
undergraduate research in computational physics.

Each instructor must decide how much computation to include, which
computational topics to cover, and how computation will be used in
the course.  These decisions must be based on a wide variety of
factors, including the computational skills of the students, the
computational background of the instructor, the computational
resources (both hardware and software) available, and time and
content coverage constraints.  Nonetheless, we feel that it is
both possible and important to include computation somewhere in
the physics curriculum.  We hope that this article provides some
guidance for those who wish to include computation in the
curriculum, but cannot create a dedicated course or program in
computational physics.

\begin{acknowledgments}
One of us (JH) wishes to thank David M. Cook for his inspiring
workshop dealing with computational physics and which he attended.
He also wishes to thank Wolfgang Christian and Francisco Esquembre
for their invaluable help in developing OSP Java applications
following OSP workshops held by Wolfgang Christian, Mario Belloni,
and Anne Cox.
\end{acknowledgments}


\begin{thebibliography}{5}

\bibitem{Grayson}Diane Grayson,``Rethinking the Content of Physics Courses,'' Physics Today {\bf
59}, No.2, 31--36 (2006).

\bibitem{Fuller}Robert G Fuller,``Numerical Computations in US undergraduate
Physics Courses,'' Computing in Science \& Eng. {\bf 8}, No.5,
16--21 (2006).

\bibitem{programs} Rubin Landau, ``Computational Physics:  A Better Model for Physics Education?,'' Computing in Science \& Eng. {\bf 8}, No. 5, 22--30 (2006);  Marty Johnston, ``Implementing Curricular Change,'' Computing in Science \& Eng. {\bf 8}, No. 5, 32--37 (2006);  Jaime R. Taylor and B. Alex King III, ``Using Computational Methods to Reinvigorate an Undergraduate Physics Curriculum,'' Computing in Science \& Eng. {\bf 8}, No. 5, 38--43 (2006);  David M. Cook, \textsl{Computation in the Lawrence Physics Curriculum} (Department of Physics, Lawrence University, Appleton, WI, 2006);  H. Gould, ``Computational physics and the undergraduate curriculum,'' Computer Physics Communications {\bf 127}, 6--10 (2000).

\bibitem{comptexts} Harvey Gould, Jan Tobochnik, and Wolfgang Christian,\textsl{An
Introduction to Computer Simulation Methods: Applications to
physical systems} (Addison-Wesley, New York, NY, 2007),3rd. ed.;
 Rubin H. Landau and Manuel J. Paez,\textsl{Computational Physics:
problem solving with computers} (John Wiley\& Sons, New York, NY,
1997).;  Steven E. Kunin and Dawn C.
Meredith,\textsl{Computational Physics: Fortran
Version}(Addison-Wesley, Reading MA, 1990);  Samuel S. M.
Wong,\textsl{Computational Methods in Physics \& Engineering}
(Prentice Hall, Englewood Clifts, NJ, 1992);  Alejandro L.
Garcia,\textsl{Numerical Methods for Physics} (Prentice Hall,
Upper Saddle River, NJ, 2000), 2nd. ed.;  Nicholas J. Giordano and
Hisao Nakanishi,\textsl{Computational Physics} (Pearson, Prentice
Hall, Upper Saddle River, NJ, 2006), 2nd. ed.;  Paul L.
DeVries,\textsl{A First Course in Computational Physics} (John
Wiley, New York, NY, 1994);  David M. Cook,\textsl{Computation and
Problem Solving in Undergraduate Physics \& Engineering}
(Department of Physics, Lawrence University, Appleton, WI, 2003).

\bibitem{Roos}Kelly R. Roos,``An Incremental Approach to Computational Physics Education,'' Computing in Science \& Eng. {\bf
8}, No.5, 44--50 (2006).
\bibitem{3M} Norman Chonacky and David Winch, ``3Ms For Instruction:  Reviews of Maple, Mathematica, and Matlab,'' Computing in Science \& Eng. {\bf 7}, No. 3, 7--13 (2005); Norman Chonacky and David Winch, ``3Ms For Instruction, Part 2: Maple, Mathematica, and Matlab,'' Computing in Science \& Eng. {\bf 7}, No. 4, 14--23 (2005).

\bibitem{OSP}See \url{www.opensourcephysics.org} as well as the OSP user's guide: Wolfgang Christian,\textsl{Open Source Physics: A user's guide with examples}
(Pearson, Addison-Wesley, San Francisco, CA, 2007).

\bibitem{SciPy}\url{www.scipy.org}.

\bibitem{EJS}See \url{www.um.es/fem/Ejs/} as well as Chapter 17 of  the OSP user's guide, Ref.\onlinecite{OSP}.

\bibitem{Belloni}Wolfgang Christian, Mario Belloni, and Douglas Brown, ``An Open-Source XML Framework for Authoring Curricular Material,'' Computing in Science \& Eng. {\bf
8}, No.5, 51--58 (2006).

\bibitem{Cromer}Alan Cromer,``Stable Solutions Using the Euler Approximation,''
Am. J Phys. {\bf 45}, 455-459 (1981).

\bibitem{texts}John R. Taylor, \textsl{Classical Mechanics} (University Science Books, Saualito, CA 2005);  Stephen T. Thornton and Jerry B. Marion,\textsl{Classical Dynamics of Particles and Systems}
(Thomson-Brooks/Cole, Belmont, CA 2004), 5th ed.; Grant R. Fowles
and George L. Cassidy,\textsl{Analytical Mechanics}
(Thomson-Brooks/Cole, Belmont, CA 2005), 7th ed.

\bibitem{Hasbun}Javier E. Hasbun,\textsl{Classical Mechanics with MATLAB Applications}
(Jones \& Bartlett, Sudbury, MA 2008), see also
\url{www.jbpub.com/catalog/0763746363/}.

\bibitem{Timberlake} Todd Timberlake, ``A computational approach to teaching conservative chaos,'' Am J. Phys. {\bf 72}, 1002--1007 (2004).

\bibitem{mscript} The matlab code used is available upon request by e-mailing the author.  OSP applications not included in Ref. \onlinecite{Hasbun} will be made
freely available here:
\url{www.westga.edu/~jhasbun/osp/osp.htm} in tandem with
the textbook release.

\bibitem{mathematica} \url{fsweb.berry.edu/academics/mans/ttimberlake/comp_phys/}.

\bibitem{bq} \url{www.bqlearning.org/}.



\end{thebibliography}
\end{document}